\documentclass{llncs} 
\usepackage{algorithm}
\usepackage{algorithmic}
\usepackage{graphicx}

\newcommand{\OPT}{\mathrm{OPT}}
\newcommand{\dist}{\mathrm{dist}}

\title{Clustering Protein Sequences Given the Approximation Stability of the Min-Sum Objective Function}

\author{Konstantin Voevodski\inst{1} \and Maria-Florina Balcan\inst{2} \and Heiko R{\"o}glin\inst{3} \and Shang-Hua Teng\inst{4} \and Yu Xia\inst{5}}

\institute{Department of Computer Science, Boston University, Boston, MA 02215, USA
\and College of Computing, Georgia Institute of Technology, Atlanta, GA 30332, USA
\and Department of Computer Science, University of Bonn, Bonn, Germany
\and Computer Science Department, University of Southern California, Los Angeles, CA 90089, USA
\and Bioinformatics Program and Department of Chemistry, Boston University, Boston, MA 02215, USA}

\begin{document}

\maketitle

\begin{abstract}
We study the problem of efficiently clustering protein sequences in a limited information setting.  We assume that we do not know the distances between the sequences in advance, and must query them during the execution of the algorithm.  Our goal is to find an accurate clustering using few queries.  We model the problem as a point set $S$ with an unknown metric $d$ on $S$, and assume that we have access to \emph{one versus all} distance queries that given a point $s \in S$ return the distances between $s$ and all other points.  Our one versus all query represents an efficient sequence database search program such as BLAST, which compares an input sequence to an entire data set.  Given a natural assumption about the approximation stability of the \emph{min-sum} objective function for clustering, we design a provably accurate clustering algorithm that uses few one versus all queries.  In our empirical study we show that our method compares favorably to well-established clustering algorithms when we compare computationally derived clusterings to gold-standard manual classifications.
\end{abstract}

\section{Introduction}

	Biology is an information-driven science, and the size of available data continues to expand at a remarkable rate.  The growth of biological sequence databases has been particularly impressive.  For example, the size of GenBank, a biological sequence repository, has doubled every 18 months from 1982 to 2007. It has become important to develop computational techniques that can handle such large amounts of data.  Clustering is very useful for exploring relationships between protein sequences. However, most clustering algorithms require distances between all pairs of points as input, which is infeasible to obtain for very large protein sequence data sets.  Even with a \emph{one versus all} distance query such as BLAST (Basic Local Alignment Search Tool) \cite{blast}, which efficiently compares a sequence to an entire database of sequences, it may not be possible to use it $n$ times to construct the entire pairwise distance matrix, where $n$ is the size of the data set.  In this work we present a clustering algorithm that gives an accurate clustering using only $O(k \log k)$ queries, where $k$ is the number of clusters.

We analyze the correctness of our algorithm under a natural assumption about the data, namely the $(c,\epsilon)$ approximation stability property of \cite{bbg}.  Balcan et al. assume that there is some relevant ``target'' clustering $C_{T}$, and optimizing a particular objective function for clustering (such as min-sum) gives clusterings that are structurally close to $C_{T}$.  More precisely, they assume that any $c$-approximation of the objective is $\epsilon$-close to $C_{T}$, where the distance between two clusterings is the fraction of misclassified points under the optimum matching between the two sets of clusters.  Our contribution is designing an algorithm that given the $(c,\epsilon)$-property for the \emph{min-sum} objective produces an accurate clustering using only $O(k \log k)$ \emph{one versus all} distance queries, and has a runtime of $O(k \log (k) n \log (n))$.  We conduct an empirical study that compares computationally derived clusterings to those given by gold-standard classifications of protein evolutionary relatedness.  We show that our method compares favorably to well-established clustering algorithms in terms of accuracy.  Moreover, our algorithm easily scales to massive data sets that cannot be handled by traditional algorithms.

The algorithm presented here is related to the one presented in \cite{vbrtx}.  The \emph{Landmark-Clustering} algorithm presented there gives an accurate clustering if the instance satisfies the $(c,\epsilon)$-property for the $k$-median objective.  However, if the property is satisfied for the \emph{min-sum} objective the structure of the clustering instance is quite different, and the algorithm given in \cite{vbrtx} fails to find an accurate clustering in such cases.  Indeed, the analysis presented here is also quite different.  The min-sum objective is also considerably harder to approximate.  For $k$-median the best approximation guarantee is $(3+\epsilon)$ given by \cite{kMedApprox}.  For the min-sum objective when the number of clusters is arbitrary there is an $O(\delta^{-1} \log^{1+\delta}n)$-approximation algorithm with running time $n^{O(1/\delta)}$ due to \cite{bcr}.

There are also several other clustering algorithms that are applicable in our limited information setting \cite{av,ajm,sampleBasedClustering,sampleBasedClustering3}.  However, because all of these methods seek to approximate an objective function they will not produce an accurate clustering in our model if the $(c,\epsilon)$-property holds for values of $c$ for which finding a $c$-approximation is difficult.  Other than \cite{vbrtx} we are not aware of any results providing both provably accurate algorithms and strong query complexity guarantees in such a model.

\section{Preliminaries}

Given a metric space $M=(X,d)$ with point set $X$, an unknown distance function
$d$ satisfying the triangle inequality, and a set of points $S \subseteq X$, we
would like to find a $k$-clustering $C$ that partitions the points in $S$ into
$k$ sets $C_{1},\ldots,C_{k}$ by using \emph{one versus all} distance
queries.

The \emph{min-sum} objective function for clustering is to minimize \linebreak $\Phi(C) = \sum_{i=1}^{k}\sum_{x,y \in C_{i}}d(x,y)$.  We reduce the min-sum clustering problem to the related \emph{balanced k-median} problem.  The balanced $k$-median objective function seeks to minimize $\Psi(C) = \sum_{i=1}^{k}\vert C_{i} \vert \sum_{x \in C_{i}} d(x,c_{i})$, where $c_{i}$ is the median of cluster $C_{i}$, which is the point $y \in C_{i}$ that minimizes $\sum_{x \in C_{i}}d(x,y)$.  As pointed out in \cite{bcr}, in metric spaces the two objective functions are related to within a factor of 2: $\Psi(C)/2 \le \Phi(C) \le \Psi(C)$.  For any objective function $\Omega$ we use $\OPT_{\Omega}$ to denote its optimum value.

In our analysis we assume that $S$ satisfies the $(c,\epsilon)$-property of \cite{bbg} for the min-sum and balanced $k$-median objective functions.  To formalize the $(c,\epsilon)$-property we need to define a notion of distance
between two $k$-clusterings $C = \lbrace C_{1},\ldots,C_{k} \rbrace$ and
$C' = \lbrace C'_{1},\ldots,C'_{k} \rbrace$. As in \cite{bbg}, we define the
distance between $C$ and $C'$ as the fraction of points on which they disagree
under the optimal matching of clusters in $C$ to clusters in $C'$:
\begin{displaymath}
   \dist(C,C') = \min_{\sigma \in S_{k}} \frac{1}{n} \sum_{i=1}^{k} \vert C_{i} - C_{\sigma(i)}' \vert,
\end{displaymath}
where $S_{k}$ is the set of bijections $\sigma\colon \lbrace 1,\ldots,k  \rbrace \rightarrow \lbrace 1,\ldots,k  \rbrace$.  Two clusterings $C$ and $C'$ are said to be \emph{$\epsilon$-close}
if $\dist(C,C') < \epsilon$.

We assume that there exists some unknown relevant ``target'' clustering $C_{T}$ and given a proposed clustering $C$ we define the error of $C$ with respect to $C_{T}$ as $\dist(C,C_{T})$. Our goal is to find a clustering of low error.  The $(c,\epsilon)$ approximation stability property is defined as follows.
\begin{definition}
We say that the instance $(S,d)$ satisfies the $(c,\epsilon)$-property for objective function $\Omega$ with respect to the target clustering $C_T$ if any
clustering of $S$ that approximates $\OPT_{\Omega}$ within a factor of $c$ is
$\epsilon$-close to $C_{T}$, that is,
$
   \Omega(C) \le c \cdot \OPT_{\Omega} \Rightarrow \dist(C,C_{T}) < \epsilon.
$
\end{definition}

We note that because any $(1+\alpha)$-approximation of the balanced $k$-median objective is a $2(1+\alpha)$-approximation of the min-sum objective, it follows that if the clustering instance satisfies the $(2(1+\alpha),\epsilon)$-property for the min-sum objective, then it satisfies the $(1+\alpha,\epsilon)$-property for balanced $k$-median.

\section{Algorithm Overview}

In this section we present a clustering algorithm that given the $(1+\alpha,\epsilon)$-property for the balanced $k$-median objective finds an accurate clustering using few distance queries.   Our algorithm is outlined in Algorithm~\ref{alg-main} (with some implementation details omitted).  We start by uniformly at random choosing $n'$ points that we call \emph{landmarks}, where $n'$ is an appropriate number.  For each landmark that we choose we use a \emph{one versus all} query to get the distances between this landmark and all other points.  These are the only distances used by our procedure.

Our algorithm then expands a ball $B_{l}$ around each landmark $l$ one point at a time.  In each iteration we check whether some ball $B_{l^{\ast}}$ passes the test in line 7.  Our test considers the size of the ball and its radius, and checks whether their product is greater than the threshold $T$.  If this is the case, we consider all balls that overlap $B_{l^{\ast}}$ on any points, and compute a cluster that contains all the points in these balls.  Points and landmarks in the cluster are then removed from further consideration.

\vspace{-0.5cm}
\begin{algorithm}[H]
\caption{Landmark-Clustering-Min-Sum($S,k,n',T$)}
\begin{algorithmic}[1]
\STATE choose a set of landmarks $L$ of size $n'$ uniformly at random from $S$;
\STATE $i=1$, $r = 0$;
\WHILE{$i \le k$}
\FOR{each $l \in L$}
\STATE $B_{l} = \{s \in S \mid d(s,l) \le r \}$;
\ENDFOR
\IF{$\exists l^{\ast} \in L \ \colon \vert B_{l^{\ast}} \vert \cdot r > T$}
\STATE $L' = \lbrace l \in L \ \colon \ B_{l} \cap B_{l^{\ast}} \ne \emptyset \rbrace$;
\STATE $C_{i} = \lbrace s \in S \ \colon \ s \in B_{l}$ and $l \in L' \rbrace$;
\STATE $i=i+1$;
\STATE remove clustered points from consideration;
\ENDIF
\STATE increment $r$ to the next relevant distance;
\ENDWHILE
\RETURN $C = \lbrace C_{1},\ldots C_{k} \rbrace$;
\end{algorithmic}
\label{alg-main}
\end{algorithm}
\vspace{-0.5cm}

A complete description of this algorithm can be found in the next section.  We now present our theoretical guarantee for Algorithm~\ref{alg-main}.

\begin{theorem}\label{thm:Main}
Given a metric space $M = (X,d)$, where $d$ is unknown, and a set of points $S$,
if the instance $(S,d)$ satisfies the $(1+\alpha,\epsilon)$-property for the balanced-$k$-median objective function, we are given the optimum objective value $\OPT$, and each cluster in the target clustering $C_{T}$ has size at least
$(6+240/\alpha)\epsilon n$, then \emph{Landmark-Clustering-Min-Sum}($S,k,n',\frac{\alpha \OPT}{40 \epsilon n}$) outputs a clustering that is $O(\epsilon/\alpha)$-close to $C_{T}$ with probability at least $1-\delta$.   The algorithm uses $n' = \frac{1}{(3+120/\alpha)\epsilon} \ln \frac{k}{\delta}$ \emph{one versus all distance queries}, and has a runtime of $O(n' n \log n)$.
\end{theorem}

We note that $n' = O(k \ln \frac{k}{\delta})$ if the sizes of the target clusters are balanced.  In addition, if we do not know the value of $\OPT$, we can still find an accurate clustering by running Algorithm~\ref{alg-main} from line 2 at most $n' n^2$ times with increasing estimates of $T$ until enough points are clustered.  It is not necessary to recompute the landmarks, so the number of distance queries that are required remains the same.  We next give some high-level intuition for how our procedures work.

\begin{figure}
\begin{center}
\includegraphics[width=45mm,height=30mm]{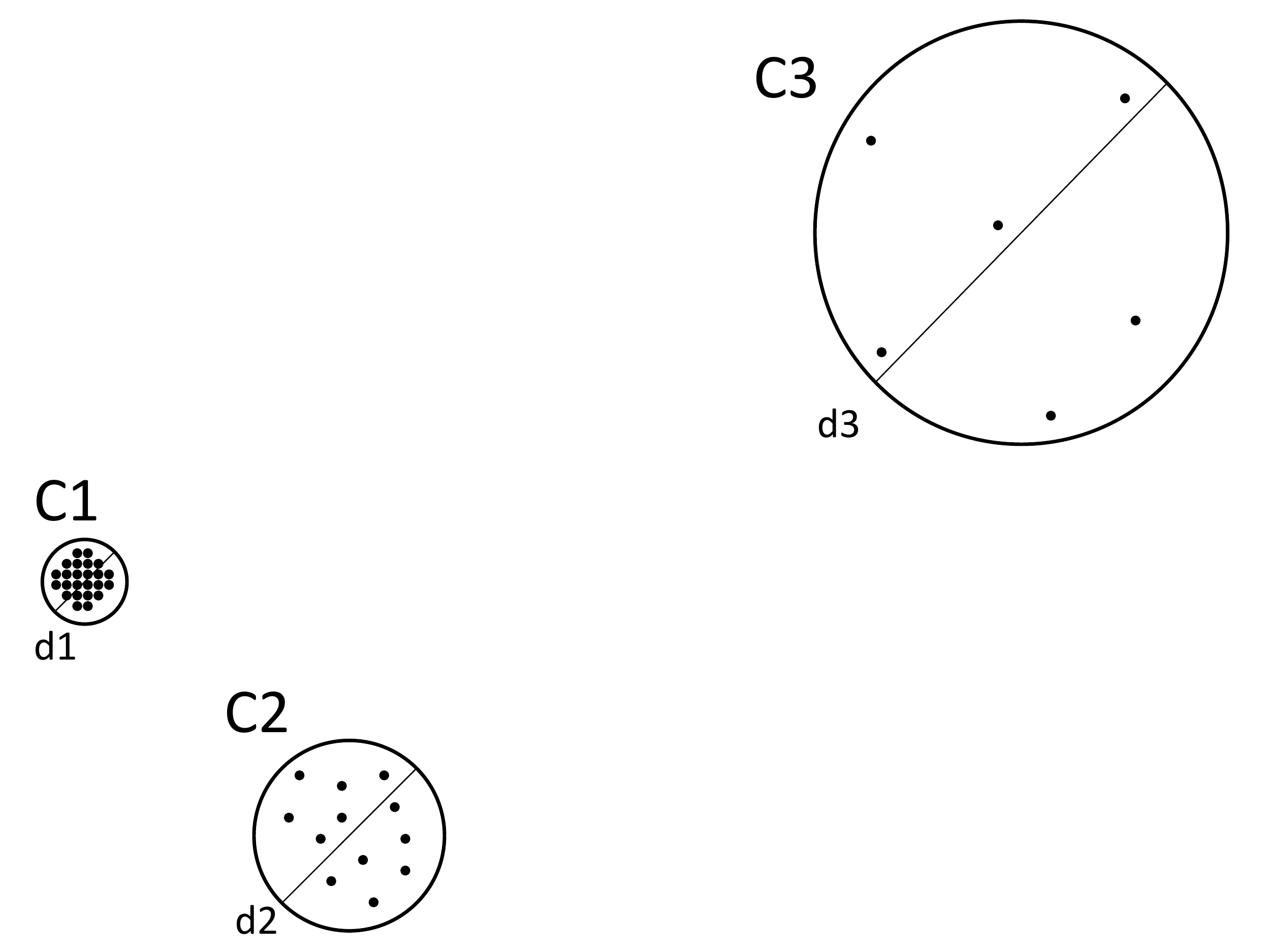}
\caption{Cluster cores $C_{1}$, $C_{2}$ and $C_{3}$ are shown with diameters $d_{1}$, $d_{2}$ and $d_{3}$, respectively.  The diameters of the cluster cores are inversely proportional to their sizes.  \label{fig:minSumStructure}}
\end{center}
\end{figure}

Given our approximation stability assumption, the target clustering must have the structure shown in Figure~\ref{fig:minSumStructure}.  Each target cluster $C_{i}$ has a ``core'' of well-separated points, where any two points in the cluster core are closer than a certain distance $d_{i}$ to each other, and any point in a different core is farther than $c d_{i}$, for some constant $c$.  Moreover, the diameters of the cluster cores are inversely proportional to the cluster sizes: there is some constant $\theta$ such that $\vert C_{i} \vert \cdot d_{i} = \theta$ for each cluster $C_{i}$.  Given this structure, it is possible to classify the points in the cluster cores correctly if we extract the smaller diameter clusters first.  In the example in Figure~\ref{fig:minSumStructure}, we can extract $C_{1}$, followed by $C_{2}$ and $C_{3}$ if we choose the threshold $T$ correctly and we have selected a landmark from each cluster core.  However, if we wait until some ball contains all of $C_{3}$, $C_{1}$ and $C_{2}$ may be merged.

\section{Algorithm Analysis}

In this section we present a formal analysis of our algorithm, and give the proof of Theorem~\ref{thm:Main}.  We first present a complete description of the algorithm.  We then describe the structure of the clustering instance that is implied by our approximation stability assumption.  We then give a general overview of our argument, which is followed by the complete proof.

\subsection{Algorithm Description}

A full description of our algorithm is given in Algorithm~\ref{alg-full}.  In order to efficiently expand a ball around each landmark, we first sort all landmark-point pairs $(l,s)$ by $d(l,s)$.  We then consider these pairs in order of increasing distance (line 7), skipping pairs where $l$ or $s$ have already been clustered; the clustered points are maintained in the set $\bar{S}$.

In each iteration we check whether some ball $B_{l^{\ast}}$ passes the test in line 19.  Our actual test, which is slightly different than the one presented earlier, considers the size of the ball and the \emph{next largest} landmark-point distance (denoted by $r_{2}$), and checks whether their product is greater than the threshold $T$.  If this is the case, we consider all balls that overlap $B_{l^{\ast}}$ on any points, and compute a cluster that contains all the points in these balls.  Points and landmarks in the cluster are then removed from further consideration by adding the clustered points to $\bar{S}$, and removing the clustered points from any ball.

Our procedure terminates once we find $k$ clusters.  If we reach the final landmark-point pair, we stop and report the remaining unclustered points as part of the same cluster (line 12).  If the algorithm terminates without partitioning all the points, we assign each remaining point to the cluster containing the closest clustered landmark.  In our analysis we show that if the clustering instance satisfies the $(1+\alpha,\epsilon)$-property for the balanced $k$-median objective function, our procedure will output exactly $k$ clusters.

The most time-consuming part of our algorithm is sorting all landmark-points pairs, which takes $O(\vert L \vert n \log n)$, where $n$ is the size of the data set and $L$ is the set of landmarks.  With a simple implementation that uses a hashed set to store the points in each ball, the total cost of computing the clusters and removing clustered points from active balls is at most $O(\vert L \vert n)$ each.  All other operations take asymptotically less time, so the overall runtime of our procedure is $O(\vert L \vert n \log n)$.

\begin{algorithm}[H]
\caption{Landmark-Clustering-Min-Sum($S,k,n',T$)}
\begin{algorithmic}[1]
\STATE choose a set of landmarks $L$ of size $n'$ uniformly at random from $S$;
\FOR{each $l \in L$}
\STATE $B_{l} = \emptyset$;
\ENDFOR
\STATE $i=1$, $\bar{S} = \emptyset$;
\WHILE{$i \le k$}
\STATE $(l,s)$ = GetNextActivePair();
\STATE $r_{1} = d(l,s)$;
\IF{($(l',s')$ = PeekNextActivePair()) $!=$ null}
\STATE $r_{2} = d(l',s')$;
\ELSE
\STATE $C_{i} = S - \bar{S}$;
\STATE break;
\ENDIF
\STATE $B_{l} = B_{l} + \lbrace s \rbrace$;
\IF{$r_{1} == r_{2}$}
\STATE continue;
\ENDIF
\WHILE{$\exists l \in L - \bar{S} \ \colon \vert B_{l} \vert > T/r_{2}$ and $i \le k$}
\STATE $l^{\ast} = \textrm{argmax}_{l \in L - \bar{S}} \vert B_{l} \vert$;
\STATE $L' = \lbrace l \in L - \bar{S} \ \colon \ B_{l} \cap B_{l^{\ast}} \ne \emptyset \rbrace$;
\STATE $C_{i} = \lbrace s \in S \ \colon \ s \in B_{l}$ and $l \in L' \rbrace$;
\FOR{each $s \in C_{i}$}
\STATE $\bar{S} = \bar{S} + \lbrace s \rbrace$;
\FOR{each $l \in L$}
\STATE $B_{l} = B_{l} - \lbrace s \rbrace$;
\ENDFOR
\ENDFOR
\STATE $i=i+1$;
\ENDWHILE
\ENDWHILE
\RETURN $C = \lbrace C_{1},\ldots C_{k} \rbrace$;
\end{algorithmic}
\label{alg-full}
\end{algorithm}

\subsection{Structure of the Clustering Instance}

We next describe the structure of the clustering instance that is implied by our approximation stability assumption.  We denote by $C^{\ast} = \lbrace C_{1}^{\ast},\ldots,C_{k}^{\ast} \rbrace$ the optimal balanced-$k$-median clustering with objective value OPT=$\Psi(C^{\ast})$.  For each cluster $C_{i}^{\ast}$, let $c_{i}^{\ast}$ be the median point in the cluster.  For $x \in C_{i}^{\ast}$, define $w(x) = \vert C_{i}^{\ast} \vert d(x,c_{i}^{\ast})$ and let $w$ = avg$_{x} w(x) = \frac{\textrm{OPT}}{n}$.  Define $w_{2}(x)$ = min$_{j \ne i} \vert C_{j}^{\ast} \vert d(x,c_{j}^{\ast})$.

It is proved in \cite{bbg} that if the instance satisfies the $(1+\alpha,\epsilon$)-property for the balanced $k$-median objective function and each cluster in $C^{\ast}$ has size at least max$(6,6/\alpha) \cdot \epsilon n$, then at most $2 \epsilon$-fraction of points $x \in S$ have $w_{2}(x) < \frac{\alpha w}{4 \epsilon}$.  In addition, by definition of the average weight $w$ at most $120\epsilon / \alpha$-fraction of points $x \in S$ have $w(x) > \frac{\alpha w}{120 \epsilon}$.

We call point $x$ \emph{good} if both $w(x) \le \frac{\alpha w}{120 \epsilon}$ and $w_{2}(x) \ge \frac{\alpha w}{4 \epsilon}$, else $x$ is called bad.  Let $X_{i}$ be the \emph{good} points in the optimal cluster $C_{i}^{\ast}$, and let $B = S \setminus \cup X_{i}$ be the bad points.

Lemma~\ref{lemma:structureBalancedKMedian}, which is similar to Lemma 14 of \cite{bbg}, proves that the optimum balanced $k$-median clustering must have the following structure:
\begin{enumerate}
\item For all $x,y$ in the same $X_{i}$, we have $d(x,y) \le \frac{\alpha w}{60 \epsilon \vert C_{i}^{\ast} \vert}$.
\item For $x \in X_{i}$ and $y \in X_{j \ne i}$, $d(x,y) > \frac{\alpha w}{5 \epsilon}/ \min(\vert C_{i}^{\ast} \vert, \vert C_{j}^{\ast} \vert)$.
\item The number of bad points is at most $b=(2+120/\alpha)\epsilon n$.
\end{enumerate}

\subsection{Proof of Theorem~\ref{thm:Main}}

Our algorithm expands a ball around each landmark, one point at a time, until some ball is large enough.  We use $r_{1}$ to refer to the current radius of the balls, and $r_{2}$ to refer to the next relevant radius (next largest landmark-point distance).  To pass the test in line 19, a ball  must satisfy $\vert B_{l} \vert > T / r_{2}$.  We choose $T$ such that by the time a ball satisfies the conditional, it must overlap some good set $X_{i}$.  Moreover, at this time the radius must be large enough for $X_{i}$ to be entirely contained in some ball; $X_{i}$ will therefore be part of the cluster computed in line 22.  However, the radius is too small for a single ball to overlap different good sets and for two balls overlapping different good sets to share any points.  Therefore the computed cluster cannot contain points from any other good set.  Points and landmarks in the cluster are then removed from further consideration.  The same argument can then be applied again to show that each cluster output by the algorithm entirely contains a single good set.  Thus the clustering output by the algorithm agrees with $C^{\ast}$ on all the good points, so it must be closer than $b + \epsilon = O(\epsilon/\alpha)$ to $C_{T}$.  A more detailed argument is given below.

\begin{proof}
Since each cluster in the target clustering has more than
$(6+240/\alpha)\epsilon n$ points, and the optimal balanced-$k$-median clustering $C^{\ast}$ can differ from the target clustering by fewer than $\epsilon n$ points, each cluster in $C^{\ast}$ must have more than $(5+240/\alpha)\epsilon n$ points.  Moreover, by Lemma~\ref{lemma:structureBalancedKMedian} we may have at most $(2+120/\alpha)\epsilon n$ bad points, and hence each $\vert X_{i} \vert = \vert C_{i}^{\ast} \setminus B \vert > (3+120/\alpha)\epsilon n \ge (2+120/\alpha)\epsilon n + 2 = b + 2$.  We will use $s_{\min}$ to refer to the $(3+120/\alpha)\epsilon n$ quantity.

Our argument assumes that we have chosen at least one landmark from each good set $X_{i}$. Lemma~\ref{lemma:numberSamples} argues that after selecting $n' = \frac{n}{s_{\min}}\textrm{ln}\frac{k}{\delta} = \frac{1}{(3+120/\alpha)\epsilon}\textrm{ln}\frac{k}{\delta}$ landmarks the probability of this happening is at least $1-\delta$.  Moreover, if the target clusters are balanced in size: $\max_{C \in C_{T}} \vert C \vert / \min_{C \in C_{T}} \vert C \vert < c$ for some constant $c$, because the size of each good set is at least half the size of the corresponding target cluster, it must be the case that $2s_{\min}c \cdot k \ge n$, so $n/s_{\min} = O(k)$.

Suppose that we order the clusters of $C^{\ast}$ such that $\vert C_{1}^{\ast} \vert \ge \vert C_{2}^{\ast} \vert \ge \ldots \vert C_{k}^{\ast} \vert$, and let $n_{i} = \vert C_{i}^{\ast} \vert$.  Define $d_{i} = \frac{\alpha w}{60 \epsilon \vert C_{i}^{\ast} \vert}$ and recall that $\max_{x,y \in X_{i}} d(x,y) \le d_{i}$.   Note that because there is a landmark in each good set $X_{i}$, for radius $r \ge d_{i}$ there exists some ball containing all of $X_{i}$.  We use $B_{l}(r)$ to denote a ball of radius $r$ around landmark $l$: $B_{l}(r): \lbrace s \in S \ \vert \ d(s,l) \le r \rbrace$.

If we apply Lemma~\ref{lemma:noOverlap} with all the clusters in $C^{\ast}$, we can see that as long as $r \le 3d_{1}$, a ball cannot contain points from more than one good set and balls overlapping different good sets cannot share any points.  We also observe that when both $r \le 3d_{1}$ and $r < d_{i}$ are true, a ball $B_{l}(r)$ containing points from $X_{i}$ does not satisfy $\vert B_{l}(r) \vert \ge T/r$.  For $r \le 3d_{1}$ a ball cannot contain points from different good sets; therefore any ball containing points from $X_{i}$ has size at most $\vert C_{i}^{\ast} \vert + b < \frac{3n_{i}}{2}$.  In addition, for $r < d_{i}$ the size bound $T/r > T / d_{i} = \frac{\alpha w}{40 \epsilon} / \frac{\alpha w}{60 \epsilon \vert C_{i}^{\ast} \vert} = \frac{3 n_{i}}{2}$.  Therefore for these values of $r$ any ball containing points from $X_{i}$ is too small to satisfy the conditional.

Finally, we observe that for $r = 3d_{1}$ some ball $B_{l}(r)$ containing all of $X_{1}$ does satisfy $\vert B_{l}(r) \vert \ge T/r$.  Clearly, for $r = 3d_{1}$ there is some ball containing all of $X_{1}$, which must have size at least $\vert C_{1}^{\ast} \vert - b \ge n_{1}/2$.  For $r = 3d_{1}$ the size bound $T/r = n_{1}/2$, so this ball is large enough to satisfy this conditional.  Moreover, for $r \le 3d_{1}$ the size bound $T/r \ge n_{1}/2$.  Therefore a ball containing only bad points cannot pass our test for $r \le 3d_{1}$ because the number of bad points is at most $b < n_{1}/2$.

Consider the smallest radius $r^{\ast}$ for which some ball $B_{l^{\ast}}(r^{\ast})$ satisfies $\vert B_{l^{\ast}}(r^{\ast}) \vert \ge T / r^{\ast}$.  It must be the case that $r^{\ast} \le 3d_{1}$, and $B_{l^{\ast}}$ overlaps with some good set $X_{i}$ because we cannot have a ball containing only bad points for $r^{\ast} \le 3d_{1}$.  Moreover, by our previous argument because $B_{l^{\ast}}$ contains points from $X_{i}$, it must be the case that $r^{\ast} \ge d_{i}$, and therefore some ball contains all the points in $X_{i}$.  Consider a cluster $\hat{C}$ of all the points in balls that overlap $B_{l^{\ast}}$: $\hat{C} = \lbrace s \in S \ \vert \ s \in B_{l}$ and $B_{l} \cap B_{l^{\ast}} \ne \emptyset \rbrace$, which must include all the points in $X_{i}$.  In addition, $B_{l^{\ast}}$ cannot share any points with balls that overlap other good sets because $r^{\ast} \le 3d_{1}$, therefore $\hat{C}$ does not contain points from any other good set.  Therefore the cluster $\hat{C}$ entirely contains some good set and no points from any other good set.

These facts suggest the following conceptual algorithm for finding a clustering that classifies all the good points correctly: increment $r$ until some ball satisfies $\vert B_{l}(r) \vert \ge T / r$, compute the cluster containing all points in balls that overlap $B_{l}(r)$, remove these points, and repeat until we find $k$ clusters.  We can argue that each cluster output by the algorithm entirely contains some good set and no points from any other good set.  Each time we can consider the clusters $C \subseteq C^{\ast}$ whose good sets have not yet been output, order them by size, and apply Lemma~\ref{lemma:noOverlap} with $C$ to argue that while $r \le 3d_{1}$ the radius is too small for the computed cluster to overlap any of the remaining good sets.  As before, we can argue that by the time we reach $3d_{1}$ we must output some cluster.  In addition, when $r \le 3d_{1}$ we cannot output a cluster of only bad points and whenever we output a cluster overlapping some good set $X_{i}$, it must be the case that $r \ge d_{i}$.  Therefore each computed cluster must entirely contain some good set and no points from any other good set.  If there are any unclustered points upon the completion of the algorithm, we can assign the remaining points to any cluster.  Still, we are able to classify all the good points correctly, so the reported clustering must be closer than $b + \dist(C^{\ast},C_{T}) < b + \epsilon = O(\epsilon/\alpha)$ to $C_{T}$.

It suffices to show that even though our algorithm only considers discrete values of $r$ corresponding to landmark-point distances, the output of our procedure exactly matches the output of the conceptual algorithm described above.  Consider the smallest (continuous) radius $r^{\ast}$ for which some ball $B_{l_{1}}(r^{\ast})$ satisfies $\vert B_{l_{1}}(r^{\ast}) \vert \ge T / r^{\ast}$.  We use $d_{real}$ to refer to the largest landmark-point distance such that $d_{real} \le r^{\ast}$.  Clearly, by the time our algorithm reaches $r_{1} = d_{real}$ it must be the case that $B_{l_{1}}$ passes the test on line 19: $\vert B_{l_{1}} \vert > T / r_{2}$, and this test is not passed by any ball at any prior time.  Moreover, $B_{l_{1}}$ must be the largest ball passing our test at this point because if there is another ball $B_{l_{2}}$ that also satisfies our test when $r_{1} = d_{real}$ it must be the case that $\vert B_{l_{1}} \vert > \vert B_{l_{2}} \vert$ because $B_{l_{1}}$ satisfies $\vert B_{l_{1}}(r) \vert \ge T / r$ for a smaller $r$.  Finally because there are no landmark-point pairs $(l,s)$ with $r_{1} < d(l,s) < r_{2}$, $B_{l}(r_{1}) = B_{l}(r^{\ast})$ for each landmark $l \in L$.  Therefore the cluster that we compute on line 22 for $B_{l_{1}}(r_{1})$ is equivalent to the cluster the conceptual algorithm computes for $B_{l_{1}}(r^{\ast})$.  We can repeat this argument for each cluster output by the conceptual algorithm, showing that Algorithm~\ref{alg-full} finds exactly the same clustering.

We note that when there is only one good set left the test in line 19 may not be satisfied anymore if $3d_{1} \ge \max_{x,y \in S} d(x,y)$, where $d_{1}$ is the diameter of the remaining good set.  However, in this case if we exhaust all landmark-points pairs we report the remaining points as part of a single cluster (line 12), which must contain the remaining good set, and possibly some additional bad points that we consider misclassified anyway.

With a simple implementation that uses a hashed set to keep track of the points in each ball, the runtime of our procedure is $O(\vert L \vert n \log n)$, which is given by the time necessary to sort all landmark-point pairs by distance.  All other operations take asymptotically less time.  In particular, over the entire run of the algorithm, the cost of computing the clusters in lines 21-22 is at most $O(n \vert L \vert)$, and the cost of removing clustered points from active balls in lines 23-28 is also at most $O(n \vert L \vert$). \qed

\end{proof}

\begin{theorem}\label{lemma:unknownW}
If we are not given the optimum objective value $w$, then we can still find a clustering that is $O(\epsilon/\alpha)$-close to $C_{T}$ with probability at least $1-\delta$ by running Landmark-Clustering-Min-Sum at most $n'n^2$ times with the same set of landmarks, where the number of landmarks $n' = \frac{1}{(3+120/\alpha)\epsilon} \ln \frac{k}{\delta}$ as before.
\end{theorem}

\begin{proof}

If we are not given the value of $w$ then we have to estimate the threshold parameter $T$ for deciding when a cluster develops.  Let us use $T^{\ast}$ to refer to its correct value ($T^{\ast} = \frac{\alpha w}{40 \epsilon}$).  We first note that there are at most $n \cdot n \vert L \vert$ relevant values of $T$ to try, where $L$ is the set of landmarks.  Our test in line 19 checks whether the product of a ball size and a ball radius is larger than $T$, and there are only $n$ possible ball sizes and $\vert L \vert n$ possible values of a ball radius.

Suppose that we choose a set of landmarks $L$, $\vert L \vert = n'$, as before. We then compute all $n' n^2$ relevant values of $T$ and order them in ascending order: $T_{i} \le T_{i+1}$ for $1 \le i < n' n^2$.  Then we repeatedly execute Algorithm~\ref{alg-full} starting on line 2 with increasing estimates of $T$.  Note that this is equivalent to trying all continuous values of $T$ in ascending order because the execution of the algorithm does not change for any $T'$ such that $T_{i} \le T' < T_{i+1}$.  In other words, when $T_{i} \le T' < T_{i+1}$, the algorithm will give the same exact answer for $T_{i}$ as it would for $T'$.

Our procedure stops the first time we cluster at least $n-b$ points, where $b$ is the maximum number of bad points.  We give an argument that this gives an accurate clustering with an additional error of $b$.

As before, we assume that we have selected at least one landmark from each good set, which happens with probability at least $1-\delta$.  Clearly, if we choose the right threshold $T^{\ast}$ the algorithm must cluster at least $n-b$ points because the clustering will contain all the good points.  Therefore the first time the algorithm clusters at least $n-b$ points for some estimated threshold $T$, it must be the case that $T \le T^{\ast}$.  Lemma~\ref{lemma:correctEstimate} argues that if $T \le T^{\ast}$ and the number of clustered points is at least $n-b$, then the reported partition must be a $k$-clustering that contains a distinct good set in each cluster.  This clustering may exclude up to $b$ points, all of which may be good points.  Still, if we arbitrarily assign the remaining points we will get a clustering that is closer than $2b + \epsilon = O(\epsilon / \alpha)$ to $C_{T}$. \qed

\end{proof}

\begin{lemma}\label{lemma:structureBalancedKMedian}
If the balanced $k$-median instance satisfies the $(1+\alpha,\epsilon$)-property and each cluster in $C^{\ast}$ has size at least $\max(6,6/\alpha) \cdot \epsilon n$ we have:
\begin{enumerate}
\item For all $x,y$ in the same $X_{i}$, we have $d(x,y) \le \frac{\alpha w}{60 \epsilon \vert C_{i}^{\ast} \vert}$.
\item For $x \in X_{i}$ and $y \in X_{j \ne i}$, $d(x,y) > \frac{\alpha w}{5 \epsilon}/ \min(\vert C_{i}^{\ast} \vert, \vert C_{j}^{\ast} \vert)$.
\item The number of bad points is at most $b=(2+120/\alpha)\epsilon n$.
\end{enumerate}
\end{lemma}

\begin{proof}
For part 1, since $x,y \in X_{i} \subseteq C_{i}^{\ast}$ are both good, they are at distance of at most $\frac{\alpha w}{120 \epsilon \vert C_{i}^{\ast} \vert}$ to $c_{i}^{\ast}$, and hence at distance of at most $\frac{\alpha w}{60 \epsilon \vert C_{i}^{\ast} \vert}$ to each other.

For part 2 assume without loss of generality that $\vert C_{i}^{\ast} \vert \ge \vert C_{j}^{\ast} \vert$.  Both $x \in C_{i}^{\ast}$ and $y \in C_{j}^{\ast}$ are good; it follows that $d(y,c_{j}^{\ast}) \le \frac{\alpha w}{120 \epsilon \vert C_{j}^{\ast} \vert}$, and $d(x,c_{j}^{\ast}) > \frac{\alpha w}{4 \epsilon \vert C_{j}^{\ast} \vert}$ because $\vert C_{j}^{\ast} \vert d(x,c_{j}^{\ast}) \ge w_{2}(x) > \frac{\alpha w}{4 \epsilon}$.  By the triangle inequality it follows that
\begin{displaymath}
d(x,y) \ge d(x,c_{j}^{\ast}) - d(y,c_{j}^{\ast}) \ge \frac{\alpha w}{\epsilon \vert C_{j}^{\ast} \vert}(\frac{1}{4} - \frac{1}{120}) > \frac{\alpha w}{5 \epsilon}/ \min(\vert C_{i}^{\ast} \vert, \vert C_{j}^{\ast} \vert),
\end{displaymath}
where we use that $\vert C_{j}^{\ast} \vert = \min(\vert C_{i}^{\ast} \vert, \vert C_{j}^{\ast} \vert)$.

Part 3 follows from the maximum number of points that may not satisfy each of the properties of the good points and the union bound. \qed
\end{proof}

\begin{lemma}\label{lemma:numberSamples}
After selecting $\frac{n}{s}\ln\frac{k}{\delta}$ points uniformly at random, where $s$ is the size of the smallest good set, the probability that we did not choose a point from every good set is smaller than $1-\delta$.

\end{lemma}

\begin{proof}
We denote by $s_{i}$ the cardinality of $X_{i}$.  Observe that the probability of not selecting a point from some good set $X_{i}$ after $\frac{nc}{s}$ samples is $(1 - \frac{s_{i}}{n})^{\frac{nc}{s}} \le (1 - \frac{s_{i}}{n})^{\frac{nc}{s_{i}}} \le (e^{-\frac{s_{i}}{n}})^{\frac{nc}{s_{i}}} = e^{-c}$.  By the union bound the probability of not selecting a point from every good set after $\frac{nc}{s}$ samples is at most $k e^{-c}$, which is equal to $\delta$ for $c = \textrm{ln}\frac{k}{\delta}$. \qed
\end{proof}

\begin{lemma}\label{lemma:noOverlap}
Given a subset of clusters $C \subseteq C^{\ast}$, and the set of the corresponding good sets $X$, let $s_{\max}= \max_{C_{i} \in C} \vert C_{i} \vert$ be the size of the largest cluster in $C$, and $d_{\min} = \frac{\alpha w}{60 \epsilon s_{\max}}$.  Then for $r \le 3 d_{\min}$, a ball cannot overlap a good set $X_{i} \in X$ and any other good set, and a ball containing points from a good set $X_{i} \in X$ cannot share any points with a ball containing points from any other good set.
\end{lemma}

\begin{proof}

By part 2 of Lemma~\ref{lemma:structureBalancedKMedian}, for $x \in X_{i}$ and $y \in X_{j \ne i}$ we have

\begin{displaymath}
d(x,y) > \frac{\alpha w}{5 \epsilon}/ \min(\vert C_{i}^{\ast} \vert, \vert C_{j}^{\ast} \vert).
\end{displaymath}

It follows that for $x \in X_{i} \in X$ and $y \in X_{j \ne i}$ we must have $d(x,y) > \frac{\alpha w}{5 \epsilon}/ \min(\vert C_{i}^{\ast} \vert, \vert C_{j}^{\ast} \vert) \ge \frac{\alpha w}{5 \epsilon}/ \vert C_{i}^{\ast} \vert > \frac{\alpha w}{5 \epsilon} / s_{\max} = 12d_{\min}$, where we use the fact that $\vert C_{i} \vert \le s_{\max}$.  So a point in a good set in $X$ and a point in any other good set must be farther than $12d_{\min}$.

To prove the first part, consider a ball $B_{l}$ of radius $r \le 3 d_{\min}$ around landmark $l$.  In other words, $B_{l} = \lbrace s \in S
\mid d(s,l) \le r \rbrace$.  If $B_{l}$ overlaps a good set in $X_{i} \in X$ and any other good set, then it must contain a point $x \in X_{i}$ and a point $y \in X_{j \ne i}$.  It follows that $d(x,y) \le d(x,l) + d(l,y) \le 2r \le 6 d_{\min}$, giving a contradiction.

To prove the second part, consider two balls $B_{l_{1}}$ and $B_{l_{2}}$ of
radius $r \le 3 d_{\min}$ around landmarks $l_{1}$ and $l_{2}$.  Suppose $B_{l_{1}}$ and $B_{l_{2}}$ share at least one
point: $B_{l_{1}} \cap B_{l_{2}} \ne \emptyset$, and use $s^{\ast}$ to refer to
this point. It follows that the distance between any
point $x \in B_{l_{1}}$ and $y \in B_{l_{2}}$ satisfies $d(x,y) \le d(x,s^{\ast})
+ d(s^{\ast},y) \le \lbrack d(x,l_{1}) + d(l_{1},s^{\ast}) \rbrack + \lbrack
d(s^{\ast},l_{2}) + d(l_{2},y) \rbrack \le 4r \le 12 d_{\min}.$

If $B_{l_{1}}$ overlaps with $X_{i} \in X$ and $B_{l_{2}}$ overlaps with $X_{j \ne i}$, and the two balls share at least one point, there must be a pair of points $x \in X_{i}$ and $y \in X_{j \ne i}$ such that $d(x,y) \le 12 d_{\min}$, giving a contradiction.  Therefore if
$B_{l_{1}}$ overlaps with some good set $X_{i} \in X$ and $B_{l_{2}}$ overlaps with any other good set, $B_{l_{1}} \cap
B_{l_{2}} = \emptyset$. \qed

\end{proof}

\begin{lemma}\label{lemma:correctEstimate}
If $T \le T^{\ast} = \frac{\alpha w}{40 \epsilon}$ and the number of clustered points is at least $n-b$, then the clustering output by Landmark-Clustering-Min-Sum using the threshold $T$ must be a $k$-clustering that contains a distinct good set in each cluster.

\end{lemma}

\begin{proof}
Our argument considers the points that are in each cluster that is output by the algorithm.  Let us call a good set \emph{covered} if any of the clusters $C_{1},\ldots, C_{i-1}$ found so far contain points from it.  We will use $\bar{C^{\ast}}$ to refer to the clusters in $C^{\ast}$ whose good sets are not \emph{covered}. It is critical to observe that if $T \le T^{\ast}$ then if $C_{i}$ contains points from an \emph{uncovered} good set, $C_{i}$ cannot overlap with any other good set.

To see this, let us order the clusters in $\bar{C^{\ast}}$ by decreasing size: $\vert C_{1}^{\ast} \vert \ge \vert C_{2}^{\ast} \vert \ge \ldots \vert C_{j}^{\ast} \vert$, and let $n_{i} = \vert C_{i}^{\ast} \vert$.  As before, define $d_{i} = \frac{\alpha w}{60 \epsilon \vert C_{i}^{\ast} \vert}$. Applying Lemma~\ref{lemma:noOverlap} with $\bar{C^{\ast}}$ we can see that for $r \le 3d_{1}$, a ball of radius $r$ cannot overlap a good set in $\bar{C^{\ast}}$ and any other good set, and a ball containing points from a good set in $\bar{C^{\ast}}$ cannot share any points with a ball containing points from any other good set.  Because $T \le T^{\ast}$ we can also argue that by the time we reach $r=3d_{1}$ we must output some cluster.

Given this observation, it is clear that the algorithm can cover at most one new good set in each cluster that it outputs.  In addition, if a new good set is covered this cluster may not contain points from any other good set. If the algorithm is able to cluster at least $n-b$ points, it must cover every good set because the size of each good set is larger than $b$.  So it must report $k$ clusters where each cluster contains points from a distinct good set. \qed

\end{proof}

\section{Experimental Results}

We present some preliminary results of testing our \emph{Landmark-Clustering-Min-Sum} algorithm on protein sequence data.  Instead of requiring all pairwise similarities between the sequences as input, our algorithm is able to find accurate clusterings by using only a few BLAST calls.  For each data set we first build a BLAST database containing all the sequences, and then compare only some of the sequences to the entire database.  To compute the distance between two sequences, we invert the bit score corresponding to their alignment, and set the distance to infinity if no significant alignment is found.  In practice we find that this distance is almost always a metric, which is consistent with our theoretical assumptions.

In our computational experiments we use data sets created from the Pfam \cite{pfam} (version 24.0, October 2009) and SCOP \cite{scop} (version 1.75, June 2009) classification databases.  Both of these sources classify proteins by their evolutionary relatedness, therefore we can use their classifications as a ground truth to evaluate the clusterings produced by our algorithm and other methods.  These are the same data sets that were used in the \cite{vbrtx} study, therefore we also show the results of the original \emph{Landmark-Clustering} algorithm on these data, and use the same amount of distance information for both algorithms ($30k$ landmarks/queries for each data set, where $k$ is the number of clusters).  In order to run \emph{Landmark-Clustering-Min-Sum} we need to set the parameter $T$.  Because in practice we do not know its correct value, we use increasing estimates of $T$ until we cluster enough of the points in the data set; this procedure is similar to the algorithm for the case when we don't know the optimum objective value $\OPT$ and hence don't know $T$. In order to compare a computationally derived clustering to the one given by the gold-standard classification, we use the distance measure from the theoretical part of our work.

Because our Pfam data sets are so large, we cannot compute the full distance matrix, so we can only compare with methods that use a limited amount of distance information.  A natural choice is the following algorithm: uniformly at random choose a set of landmarks $L$, $\vert L \vert = d$; embed each point in a $d$-dimensional space using distances to $L$; use $k$-means clustering in this space (with distances given by the Euclidian norm).  This procedure uses exactly $d$ one versus all distance queries, so we can set $d$ equal to the number of queries used by the other algorithms.  For SCOP data sets we are able to compute the full distance matrix, so we can compare with a spectral clustering algorithm that has been shown to work very well on these data \cite{spectralClusteringProteinSeqs}.

\begin{figure}
\begin{center}
\includegraphics[width=100mm,height=70mm]{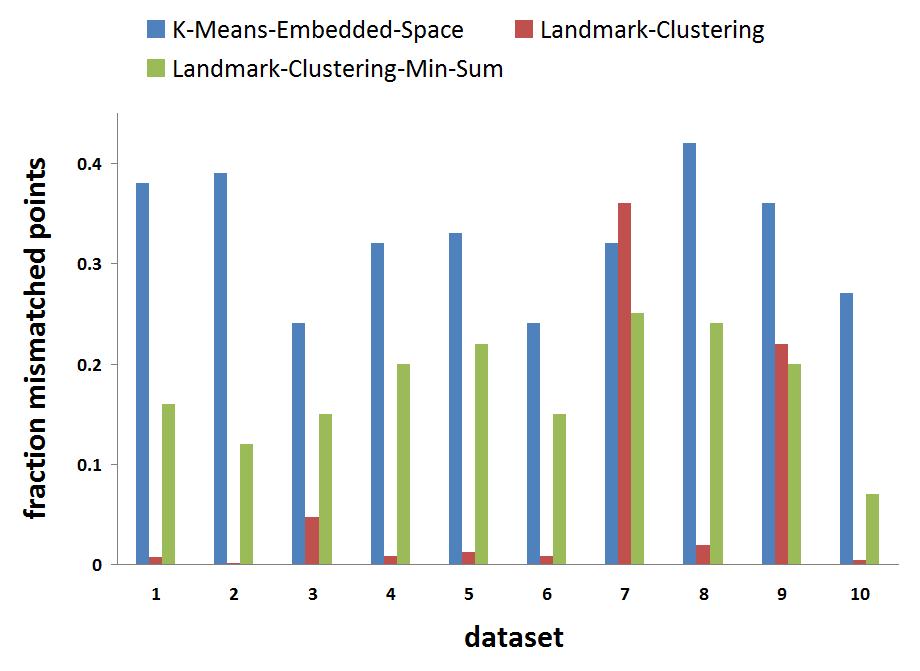}
\caption{Comparing the performance of $k$-means in the embedded space (blue), \emph{Landmark-Clustering} (red), and \emph{Landmark-Clustering-Min-Sum} (green) on 10 data sets from Pfam.  Datasets \textbf{1-10} are created by uniformly at random choosing 8 families from Pfam of size $s$, $1000 \le s \le 10000$. \label{fig:figure1}}
\end{center}
\end{figure}

From Figure~\ref{fig:figure1} we can see that \emph{Landmark-Clustering-Min-Sum} outperforms $k$-means in the embedded space on all the Pfam data sets.  However, it does not perform better than the original \emph{Landmark-Clustering} algorithm on most of these data sets.  When we investigate the structure of the ground truth clusters in these data sets, we see that the diameters of the clusters are roughly the same.  When this is the case the original algorithm will find accurate clusterings as well \cite{vbrtx}.  Still, \emph{Landmark-Clustering-Min-Sum} tends to give better results when the original algorithm does not work well (data sets 7 and 9).

\begin{figure}
\begin{center}
\includegraphics[width=100mm,height=70mm]{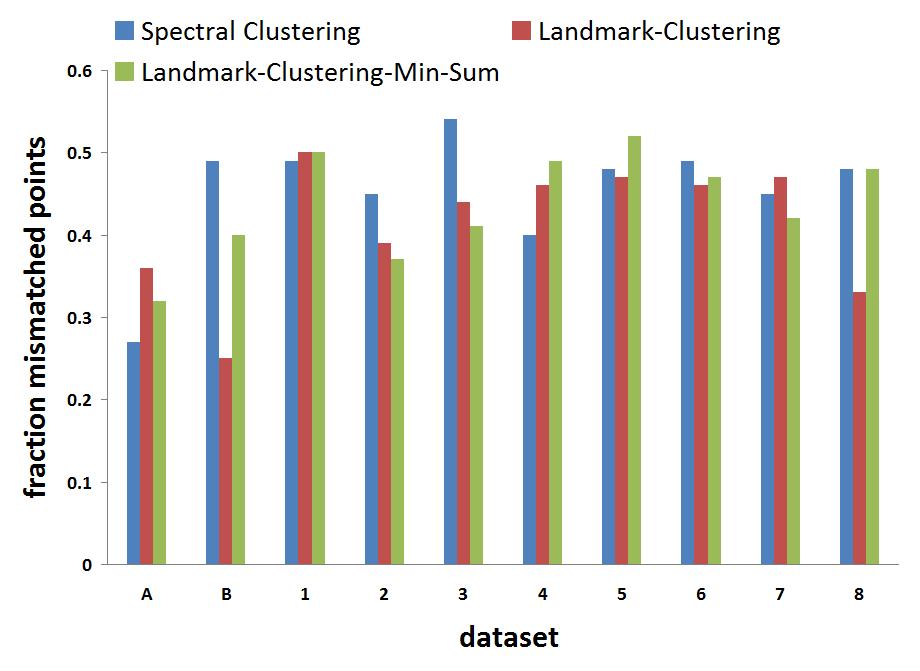}
\caption{Comparing the performance of spectral clustering (blue), \emph{Landmark-Clustering} (red), and \emph{Landmark-Clustering-Min-Sum} (green) on 10 data sets from SCOP.  Data sets \textbf{A} and \textbf{B} are the two main examples from \cite{spectralClusteringProteinSeqs}, the other data sets (\textbf{1-8}) are created by uniformly at random choosing 8 superfamilies from SCOP of size $s$, $20 \le s \le 200$.
\label{fig:figure2}}
\end{center}
\end{figure}

Figure~\ref{fig:figure2} shows the results of our computational experiments on the SCOP data sets.  We can see that the three algorithms are comparable in performance here.  These results are encouraging because the spectral clustering algorithm significantly outperforms other clustering algorithms on these data \cite{spectralClusteringProteinSeqs}. Moreover, the spectral algorithm needs the full distance matrix as input and takes much longer to run.  When we examine the structure of the SCOP data sets, we find that the diameters of the ground truth clusters vary considerably, which resembles the structure implied by our approximation stability assumption, assuming that the target clusters vary in size.  Still, most of the time the product of the cluster sizes and their diameters varies, so it does not quite look like what we assume in the theoretical part of this work.

We plan to conduct further studies to find data where clusters have different scale and there is an inverse relationship between cluster sizes and their diameters.  This may be the case for data that have many outliers, and the correct clustering groups sets of outliers together rather than assigns them to arbitrary clusters.  The algorithm presented here will consider these sets to be large diameter, small cardinality clusters.  More generally, the algorithm presented here is more robust because it will give an answer no matter what the structure of the data is like, whereas the original \emph{Landmark-Clustering} algorithm often fails to find a clustering if there are no well-defined clusters in the data.

\section{Conclusion}

We present a new algorithm that clusters protein sequences in a limited information setting.  Instead of requiring all pairwise distances between the sequences as input, we can find an accurate clustering using few BLAST calls.  We show that our algorithm produces accurate clusterings when compared to gold-standard classifications, and we expect it to work even better on data who structure more closely resembles our theoretical assumptions.

\bibliographystyle{alpha}
\bibliography{references}
\end{document}